\documentclass[preprint,showpacs,preprintnumbers,amsmath,amssymb]{revtex4}

\usepackage{graphicx}
\usepackage{dcolumn}
\usepackage{bm}

\begin{document}


\title{Pheromone Static Routing Strategy for Complex Networks}
\author{Xiang Ling$^1$}
\author{Henry Y.K. Lau$^2$}
\author{Rui Jiang$^1$}
\author{Mao-Bin Hu$^1$}\email{humaobin@ustc.edu.cn}
\affiliation{%
1.School of Engineering Science, University of Science and
Technology of China, Hefei 230026, People's Republic of China \\
2.Department of Industrial and Manufacturing Systems
Engineering, The University of Hong Kong, Pokfulam Road, Hong Kong, P.R. China\\
}%

\date{\today}

\begin{abstract}
In this paper, we adopt the concept of pheromone to generate a set
of static paths that can reach the performance of global dynamic
routing strategy [Phys. Rev. E 81, 016113(2010)]. In the test
stage, pheromone is dropped to the nodes by packets forwarded by
the global dynamic routing strategy. After that, static paths are
generated according to the density of pheromone. The output paths
can greatly improve traffic systems' overall capacity on different
network structures, including scale-free networks, small-world
networks and random graphs. Because the paths are static, the
system needs much less computational resource than the global
dynamic routing strategy.
\end{abstract}

\maketitle

\section{Introduction}

Traffic on networked systems is important for many modern
communication and transportation systems. Since the discovery of
Small-World phenomenon \cite{WS}, and Scale-Free property
\cite{BA}, it is widely proved that the topology and degree
distribution of networks have profound effects on the processes
taking place on these networks, including traffic flow
\cite{Boccaletti,Arenas,Guimera,Tadic,Echen,ZhangH,DuWB,DuWB2,DuWB3}.

Routing is the control mechanism that directs each commuter in a
networked system from its source to destination through a serial
of switching nodes. A good routing strategy can improve network
traffic efficiency without changing the underlying infrastructures
\cite{Kleinberg,YG,WWX,Ling1,Ling2}. Therefore, adopting efficient
routing strategies are often preferred by the engineers. In modern
communication and transportation systems, the shortest path
protocol is often used. However, the shortest paths often lead to
the collapse of hub nodes, which greatly reduces system
performance. To avoid this problem, some routing strategies are
proposed. One good example is the efficient routing strategy
\cite{YG}. The efficient path between any pair of nodes is defined
as the path that the sum degree of nodes is minimum, denoted as:
\begin{equation}\label{efficient}
Path=min\sum^l_{m=1}k(x_m)^\beta,
\end{equation}
where $k(x_m)$ is the degree of $m$th node along the path, $l$ is
the path length and $\beta$ is a tunable parameter. When
$\beta=1$, Eq.(\ref{efficient}) can reach a very high network
capacity, which is almost ten times larger than shortest paths.
Actually, the efficient routing strategy propels the packets to
use the low-degree nodes, which composes the peripheral of the
network. Therefore, the capacity of low-degree nodes can be
utilized and so the system's capacity can be improved.

The nodes' ability can not be fully reflected by the degree. The
best routing strategy to fully utilize the nodes' ability might be
accomplished by examine the queue length of nodes. Recently, Ling
etal proposed a global dynamic routing strategy \cite{Ling2}, in
which the path between any pair of nodes is defined as the sum of
nodes' queue length is minimum:
\begin{equation}
Path=min\sum^{l}_{m=1}[1+n(x_m)];
\end{equation}
where $n(x_m)$ is the queue length of node $x_m$. This routing
strategy selects the path along which the nodes have least traffic
loads. Thus the network capacity can be further enhanced to almost
double of the efficient routing strategy. As far as we know, the
global dynamic routing strategy can achieve the highest
performance for scale-free networks. Nevertheless, the shortcoming
of this routing protocol lies in its high computational
consumption since the paths are dynamical. This problem can be
acute when the network size is large. An approach to reach the
performance of global dynamic routing strategy with static paths
is needed.

In this brief report, we propose that by the help of ``pheromone",
this kind of static routing paths can be obtained. 
The concept of ``pheromone" has been proved to be efficient to
form a good local routing strategy \cite{Ling1}. Here it is
demonstrated that ``pheromone" can also help to generate routing
paths that can reach high traffic capacities as the global dynamic
routing strategy for scale-free, small-world and random network
infrastructures.

\section{Pheromone Static Routing Strategy}

In the process of moving, ants will release a special kind of
secretion pheromone, which helps other ants to look for paths
\cite{ACA,Bonabeau}. The pheromone released by the pioneer packets
are important for the final paths to be established. Similarly, we
suppose that some pioneer packets travel in the network by the
global dynamic paths and drop an information of ``pheromone" to
the nodes in the first period. Therefore, the traffic flow
characters of global dynamic routing strategy can be reserved.
Then the static routing paths are generated by the ``pheromone"
information. Finally, the packets are forwarded with the output
routing paths. The paths are static and the packets do not need to
drop ``pheromone" any more.

The traffic model is described as follows. Initially, the
pheromone density of each node is set to zero. The maximal queue
length of each node is assumed to be unlimited and
First-In-Fist-Out (FIFO) discipline is applied. The nodes'
delivering ability is set to $C=1$, i.e., each node can forward
only one packet to its neighboring nodes in each step.

(1) Test Stage: In this stage with preset length of $T_t$ steps,
$N_{pt}$ test packets are forwarded in the system by the global
dynamic routing strategy \cite{Ling2}. The routing paths are
update in every step. Meanwhile, each test packet waiting in the
queue will release one unit of pheromone to the current node for
each step. That is, if a test packet waits for $5$ steps in the
queue before being delivered, the pheromone density of this node
will increase by $5$ units. The unit pheromone value is set to
$\delta p=0.0001$. When a test packet arrives at its destination,
it remains in the system and is forwarded to another random
destination. After $T_t$ steps, the nodes in the system will have
different pheromone densities and the test packets are all
removed.

(2) Path Generation Stage: The pheromone static path for any pair
of nodes is generated that the sum pheromone density of nodes
along the path is minimum:
\begin{equation}\label{pheromone}
Path=min \sum^{l}_{m=1}p(x_m),
\end{equation}
where $p(x_m)$ is the pheromone density of node $x_m$.

(3) Delivery Stage: In each step, $R$ packets enter the system
with randomly chosen sources and destinations. The packets are
forwarded by the pheromone static paths. The paths do not need to
be updated. Once a packet arrives at
its destination, it is removed from the system. 
Following common practice, we adopt the order parameter
\cite{Arenas}:
\begin{eqnarray}
\eta(R)=\lim_{t\rightarrow \infty}\frac{C}{R}\frac{\langle \Delta
N_p \rangle}{\Delta t}
\end{eqnarray}
where $\Delta N_p=N_{p}(t+\Delta t)-N_{p}(t)$, $N_{p}(t)$ is the
number of packets within the network at time $t$, and
$\langle...\rangle$ indicates the average over time windows of
width $\Delta t$. With increasing packet generation rate $R$,
there will be a critical value of $R_c$ which characterizes the
phase transition from free flow ($\eta=0$) to congestion
($\eta>0$). In the free flow state, $\eta$ is around zero due to
the balance of created and removed packets. When $R$ exceeds
$R_c$, the packets accumulate in the network, and $\eta$ becomes
positive. The network's overall capacity can be measured by the
maximal generating rate $R_c$. $R_c$ can also be estimated
analytically by the maximal node betweenness in the network
\cite{Guimera}:
\begin{equation}\label{Rc}
R_c=\frac{N(N-1)}{B_{max}},
\end{equation}
where betweenness ($B$) is defined as the number of paths passing
through a node, and $N$ is the network size.

\begin{figure}
\scalebox{0.8}[0.8]{\includegraphics{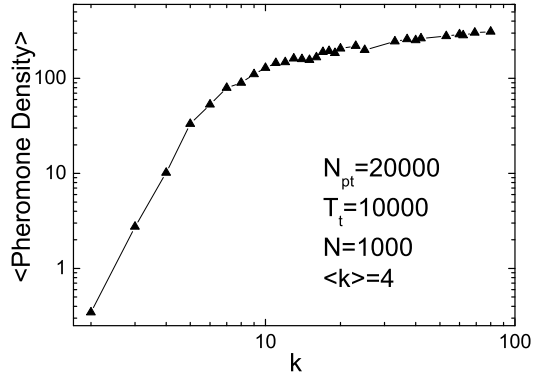}}
\caption{\label{fig1}Pheromone density vs node degree $k$ with
test packet number $N_{pt}=20000$ and testing time $T_{t}=10000$.
Network parameters are size $N=1000$ and average degree $\langle
k\rangle=4$.}
\end{figure}

\begin{figure}
\scalebox{0.9}[0.9]{\includegraphics{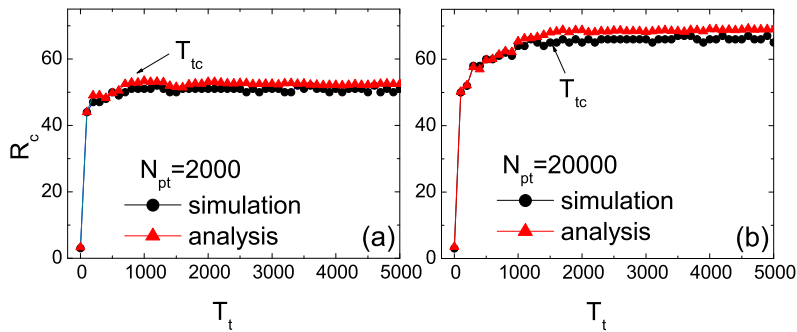}}
\caption{\label{fig2}(Color online.) Variation of system capacity
$R_c$ with the length of test period $T_t$ with $N_{pt}=2000$ (a)
and $N_{pt}=20000$ (b). Each data point is obtained by carry out
the model on one BA network realization with parameters $N=1000$
and $\langle k\rangle=4$.}
\end{figure}

We note that this model is different from the ant colony algorithm
\cite{ACA,Bonabeau} because it does not consider the evaporation
effect of pheromone and the pheromone only appears in the first
two stages. Actually, the concept of pheromone is borrowed only to
record the traffic flow trace of global dynamic routing strategy.
From the above model, one can also see that the model can be
adjusted by changing the parameters of testing time length $T_t$
and test packet number $N_{pt}$. In the following, we will show
the results on three typical network structures: the
Barab\'asi-Albert (BA) scale-free networks \cite{BA}, the
Newman-Watts small-world networks \cite{NW} and the
Erd\"os-R\'enyi (ER) random graphs \cite{ER}.


\section{Simulation and Analytical Results}

We firstly present the simulation and analytical results on BA
scale-free networks \cite{BA}.
In this model, starting from $m_0$ fully connected nodes, a new
node with $m$ links is added to the existing graph at each step
according to the preferential attachment. The probability for the
new node to be connected to an existing node $i$ is proportional
to the degree $k_i$ of the node.

Figure \ref{fig1} shows a typical distribution of pheromone as
function of node degree at the end of test stage. It approximately
follows a piecewise power law with positive exponent. This
indicates that the hub nodes are more burdened with the global
dynamic routing strategy.

\begin{figure}
\scalebox{0.9}[0.9]{\includegraphics{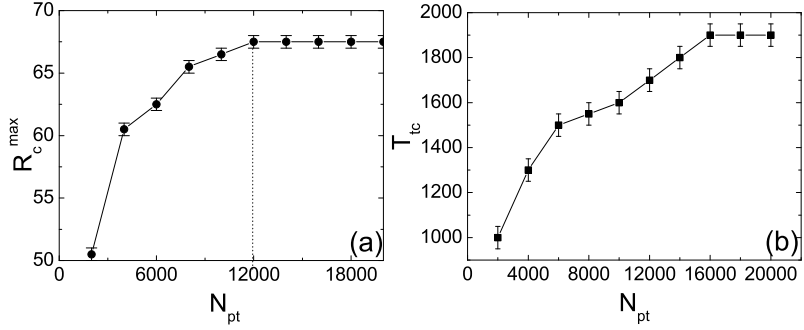}}
\caption{\label{fig3} (a) Maximal capacity $R_c^{max}$ vs $N_{pt}$
with $T_t=10000$. (b) Saturation point $T_{tc}$ vs $N_{pt}$ for
the system. Network parameters are $N=1000$ and $\langle
k\rangle=4$.}
\end{figure}

\begin{figure}
\scalebox{0.9}[0.9]{\includegraphics{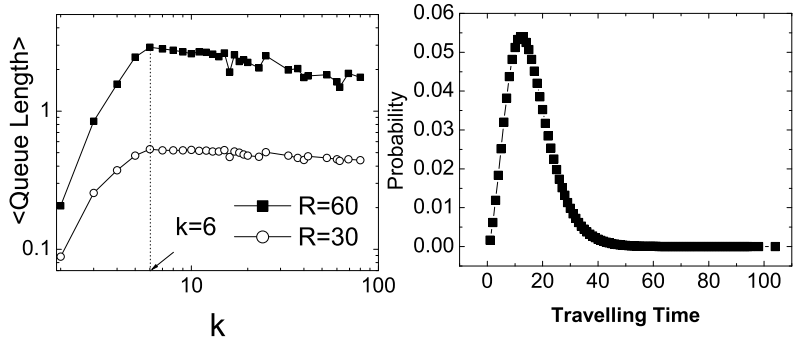}}
\caption{\label{fig4} (a) Average queue length vs node degree $k$
with the pheromone static routing strategy in free flow state. (b)
The probability distribution of travelling time in free flow state
($R=60$). Network parameters are $N=1000$ and $\langle
k\rangle=4$.}
\end{figure}

Figure \ref{fig2} shows the results of network traffic capacity
measured by $R_c$ with changing $T_t$. The analytical estimation
of $R_c$ by Eq.(\ref{Rc}) are shown with Betweenness calculated by
the pheromone static paths.
For a given $N_{pt}$, network capacity $R_c$ increases with $T_t$
until it reaches saturation at a critical point of $T_{tc}$. The
saturated network capacity $R_c^{max}$ is affected by $N_{pt}$.
When $N_{pt}=12000$, the maximum network capacity is $R_c^{max}=
68$. This is very close to that of the global dynamic routing
strategy ($R_c=71$) \cite{Ling2} and is more than double of
efficient routing strategy ($R_c=33$) \cite{YG} with the same
network parameters. One can see that the analytical results are
slightly larger than simulation results. This discrepancy might be
due to the noise in the simulation.

In Fig.\ref{fig3}(a), one can see that $R_c^{max}$ increases with
$N_{pt}$ and come to a saturate value of $68$ at $N_{pt}=12000$.
In Fig.\ref{fig3}(b), we also show the variation of critical point
$T_{tc}$ for different $N_{pt}$.


\begin{figure}
\scalebox{0.7}[0.7]{\includegraphics{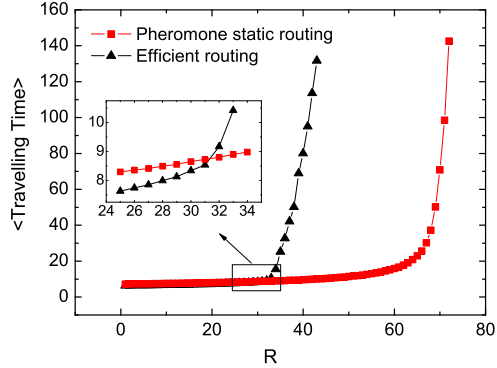}}
\caption{\label{fig5}(Color online). Average travelling time for
different $R$ with network size $N=1000$ and average degree
$\langle k\rangle=4$. When $R>R_c$, the average travelling time is
calculated by averaging the travelling time of arriving packets
during $1000$ steps of delivery.}
\end{figure}

Figure \ref{fig4}(a) shows the distribution of average queue
length as function of node degree $k$ with the pheromone static
paths. For both cases $R=60<R_c$ and $R=30<R_c$, one can see that
the traffic load distributes non-linearly with a maximum at $k=6$.
This indicates that the traffic load on hub nodes is alleviated.
This behavior is different from the situations of shortest path
and global dynamic routing strategy where the queue length follows
a power law of $n(k) \sim k^\gamma$ with exponent $\gamma>0$.
Figure \ref{fig4}(b) shows the probability distribution of
packet's travelling time in free flow state ($R=60<R_c$) with the
pheromone static paths. The travelling time is recorded as the
time that the packet spend from source to destination. It
approximately follows a Poisson distribution, which is similar to
that of global dynamic routing strategy \cite{Ling2}.

Figure \ref{fig5} shows the variation of packets' average
travelling time over packet generating rate $R$ for the efficient
routing strategy and pheromone static paths. one can see that the
average travelling time is slightly larger than that with
efficient routing strategy for $R<32$. When $R>32$, the system
enters congestion for efficient routing strategy, thus the
packet's travelling time increases rapidly. When $R>68$, the
system is also congested with pheromone static paths, thus the
average travelling time also increases.

\begin{figure}
\scalebox{0.9}[0.9]{\includegraphics{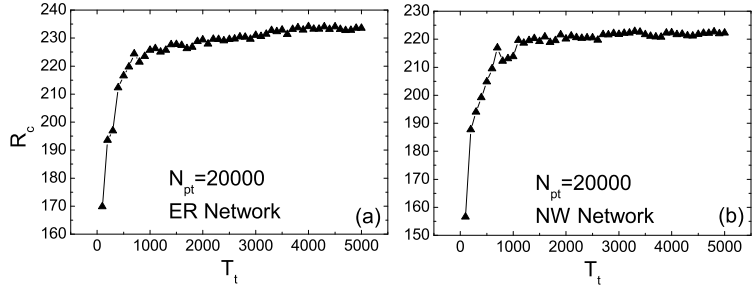}}
\caption{\label{fig6} Analytical estimation of network traffic
capacity $R_c$ vs testing time $T_t$ with $N_{pt}=20000$ for: (a)
Erdos-Renyi random networks of $N=1000$ and $\langle k\rangle=8$;
(b) Newman-Watts small-world networks of $N=1000$ and $\langle
k\rangle=8$.}
\end{figure}




Finally we present the results of network traffic capacity with
Newman-Watts (NW) small-world networks \cite{NW} and
Erd\"os-R\'enyi (ER) random graphs \cite{ER}. NW small-world
networks are generated by randomly adding long-range links to a
one-dimensional lattice with nearest and next-nearest interactions
with periodic boundary conditions \cite{NW}. ER random graphs are
generated by connecting couples of randomly selected nodes,
prohibiting multiple connections \cite{ER}. We find the model can
also achieve much higher system capacity on these networks. In
Fig.\ref{fig6}, the analytical results of network traffic capacity
measured by $R_c$ are shown for the two type of networks. The
overall capacity can reach a maximum of $R_c^{max} \approx 234$
for ER networks and $R_c^{max} \approx 222$ for NW networks. For
comparison, we find by simulation that $R_c \approx 50$ with
shortest path routing strategy and $R_c \approx 140$ with
efficient routing strategy for ER networks with the same
parameters. For NW networks, $R_c \approx 85$ with shortest path
routing strategy and $R_c \approx 155$ with efficient routing
strategy.

\section{Summary}

In summary, a set of static routing paths is proposed for the
traffic system on complex network infrastructures. Remarkably, the
paths can almost reach the high performance of global dynamic
routing strategy with a fixed routing table. It is demonstrated
that the model can perform well on different network structures,
include scale-free, small-world and random networks. For
scale-free networks, the distribution of queue length and
probability of travelling time are also investigated. This model
can help the plan and design of routing protocols of modern
communication and transportation networks.


\begin{acknowledgements}
This work is funded by the National Natural Science Foundation of
China with Nos.10872194, 11072239, the Fundamental Research Funds
for the Central Universities of China. M.-B. Hu acknowledges the
support of Massey University International Visitor Research Fund
and the Endeavour Australia Cheung Kong Research Fellowship.
\end{acknowledgements}

\end{document}